\documentclass[letterpaper,twocolumn,10pt]{article}
\usepackage{times}
\usepackage{url}
\usepackage{breakurl}
\usepackage{epsfig}
\usepackage{graphics}
\usepackage{times}
\usepackage{comment}
\usepackage{floatflt}
\usepackage{import}
\usepackage[cmex10]{amsmath}
\usepackage{amssymb}
\usepackage{listings}
\usepackage{etoolbox}
\usepackage[usenames, dvipsnames]{xcolor}
\usepackage[letterpaper,top=1.0in,bottom=1.0in,inner=1.0in,outer=1.0in]{geometry}

\usepackage[compact]{titlesec}
\usepackage[]{algorithm2e}
\usepackage{transparent}
\usepackage{color,colortbl}
\usepackage[T1]{fontenc}
\usepackage{textcomp} 
\usepackage{xspace}
\usepackage{pdfpages}
\usepackage{subcaption}
\usepackage{titling}
\usepackage[subtle=yes]{savetrees}

  {\begin{list}{$\bullet$}%
     {\setlength{\parsep}{0pt}%
      \setlength{\topsep}{0pt}%
      \setlength{\itemsep}{0pt}}}%
{\end{list}}

\newcommand{\para}[1]{\medskip\noindent{\bf #1}}

\def\compactify{\itemsep=0pt \topsep=0pt \partopsep=0pt \parsep=0pt}
\let\latexusecounter=\usecounter

\newcommand{\textred}[1]{\textcolor{red}{#1}}
\ifx\noeditingmarks\undefined
   \newcommand{\pgwrapper}[2]{\textred{#1: #2}}
\else
   \newcommand{\pgwrapper}[2]{}
\fi

\newcommand{\nop}[1]{}

\newcommand{\name}{Gringotts\xspace}

\date{}
\title{Secure Incentivization for Decentralized Content Delivery\vspace{-0.5cm}}

\author{Prateesh Goyal$^{1}$, Ravi Netravali$^{1,2}$, Mohammad Alizadeh$^{1}$, Hari Balakrishnan$^{1}$\\MIT CSAIL$^{1}$, UCLA$^{2}$}

\begin{document}
\maketitle
\begin{sloppypar}
\begin{abstract}

Prior research has proposed technical solutions to use peer-to-peer (P2P) content delivery to serve Internet video, showing that it can reduce costs to content providers. Yet, such methods have not become widespread except for a few niche instances. An important challenge is incentivization: what tangible benefits does P2P content delivery offer users who bring resources to the table? In this paper, we ask whether monetary incentives can help attract peers in P2P content delivery systems. We commissioned a professional survey of people around the United States to answer several relevant questions. We found that 51\% of the 876 respondents---substantially larger than our expectations---answered ``yes'' to whether they would participate for suitable financial incentives.
Encouraged by the results of the survey, we propose \name{}, a system to structure incentives and  securely incorporate P2P delivery into content delivery systems. \name{} provides a novel Proof of Delivery mechanism that allows content providers to verify correct delivery of their files, and shows how to use cryptocurrency to pay peers while guarding against liars and Sybil attacks. 

\end{abstract}
\section{Introduction}
\label{s:intro}

Video streams constitute over 70\% of global Internet traffic~\cite{ciscovideo,sandvine}. Most video is delivered to users today via Content Distribution Networks (CDNs) like Akamai and CloudFlare. Although CDN demand has nearly doubled since 2016~\cite{ciscovideo}, they are too expensive for many content providers~\cite{cdntrends}.

To combat these high prices, there have been many proposals to have peers cache and stream videos to each other, either as supplements to existing CDNs~\cite{netsessions,iqiyi,livesky}, or as decentralized P2P systems~\cite{peer5,dandelion}. For example, peers within an Internet Service Provider (ISP) in a city could stream videos directly to one another, avoiding expensive Internet paths. By serving content using bandwidth and storage resources that would otherwise go unused, such systems could significantly reduce costs for content providers and CDN operators.

These proposals have not seen significant adoption in practice because of a lack of sufficient participation~\cite{p2psurvey,netsessions}. Even forms of incentivization like virtual tokens and in-service perks (e.g., traffic prioritization) have failed to attract enough peers~\cite{dandelion, antfarm,bittyrant}. We believe that the steady increases in uplink bandwidth and storage capacity on laptops~\cite{speedtest} motivate revisiting P2P content delivery, but with a focus on secure payments as an incentive mechanism. 


We ask two questions. First, would users be willing to participate in such a system if they were incentivized with monetary payments? To answer this question, we commissioned a consumer survey to understand user concerns with respect to participation in P2P content delivery (\S\ref{s:userstudy}). Our key finding is that 51\% of the 876 users would participate. We were pleasantly surprised by this percentage, which is larger than what we expected. Those who would not participate were primarily concerned with device security, content liability, and impacts on device performance. To our knowledge, the results we report are the first published findings on this question.


Second, how can payments for monetary incentivization be done securely? This is challenging for several reasons. First, a content provider cannot be trusted to honor payments, and peers may be disinclined to share payment credentials with certain content providers. 
Second, existing centralized payment systems (e.g., Paypal) are not designed to support the large number of small transactions that would be needed for P2P content delivery. This is particularly challenging because peers and content providers can span geographic boundaries that impose foreign transaction/exchange fees. Further, these issues are more pronounced for small content providers, both due to financial limitations and difficulties in convincing peers of faithful payments.

One approach to address these challenges is to use a central authority that everyone trusts (e.g., a bank) to enforce content delivery payments. An alternative approach is to eliminate the use of a central authority (and the associated trust requirements), and to instead provide the above guarantees in a completely \textit{decentralized} manner. With either approach, content providers must be assured that delivery is happening properly. Our primary contribution is a lightweight solution to this problem in which content providers, clients, and peers, collectively produce a {\em Proof of Delivery Chain (PoDC) } that serves as a proof for the delivery of a file from a set of peers to a client. PoDCs are unforgeable and tamper-proof: neither the peers nor the content providers can manipulate them to affect payments.



We present the design of \name{}, a system that applies PoDC to decentralized P2P content delivery. Payments in \name{} are made using a cryptocurrency. Cryptocurrencies naturally address some of the aforementioned challenges, and our consumer survey revealed that 27\% of users are already willing to accept cryptocurrency payments (with 40\% unsure due to lack of familiarity). In \name{}, peers are ensured payments by broadcasting PoDCs on a blockchain. \name uses probabilistic payments to limit blockchain transaction overheads without compromising security. \name is also robust to various forms of collusion (e.g., clients and content providers, clients and peers) and Sybil attacks with fake clients or peers.

\section{Consumer Survey}
\label{s:userstudy}

To understand the expectations and requirements for user participation in a peer-to-peer content delivery service, we commissioned a third-party professional organization to undertake a consumer survey. They survey garnered 876 responses from around the United States. The respondents ranged in age, with 95\% between the ages of 18-60; 52\% of the participants were female; annual household incomes were distributed between \$10K-200K+, with 45\% between \$25K-\$100K. Each participant was asked a set of 11 questions, relating to payments, resource availability, and participation concerns.

Our key findings are:
\begin{itemize}
\itemsep-0.25em
\item 51\% of users said they would participate.
\item Of those who would participate, 70\% expect to earn no more than 50\% of their monthly Internet bill.
\item 27\% of users are willing to accept payment in the form of cryptocurrency; 40\% were unsure about cryptocurrency payments, while 33\% were unwilling.
\item The largest concerns for users who would not participate were security and privacy concerns (83\%), liability concerns over content (50\%), and impact on device performance (47\%).
\end{itemize}

\begin{figure}
    \centering
    \begin{subfigure}[b]{0.23\textwidth}
        \includegraphics[width=\textwidth]{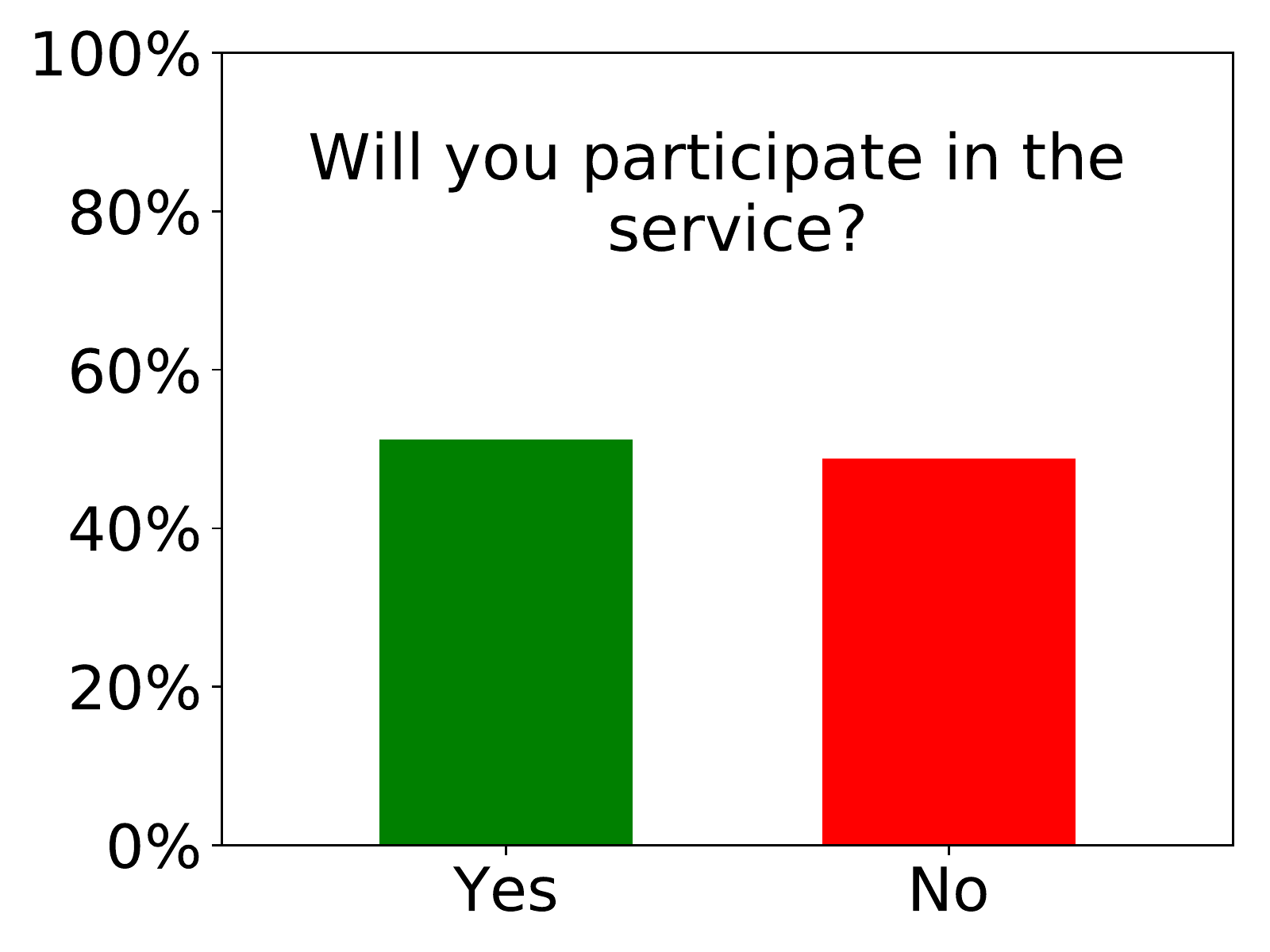}
        \vspace{-0.7cm}
        \caption{}
        \label{fig:key_finding:participation}
    \end{subfigure}
    \begin{subfigure}[b]{0.23\textwidth}
        \includegraphics[width=\textwidth]{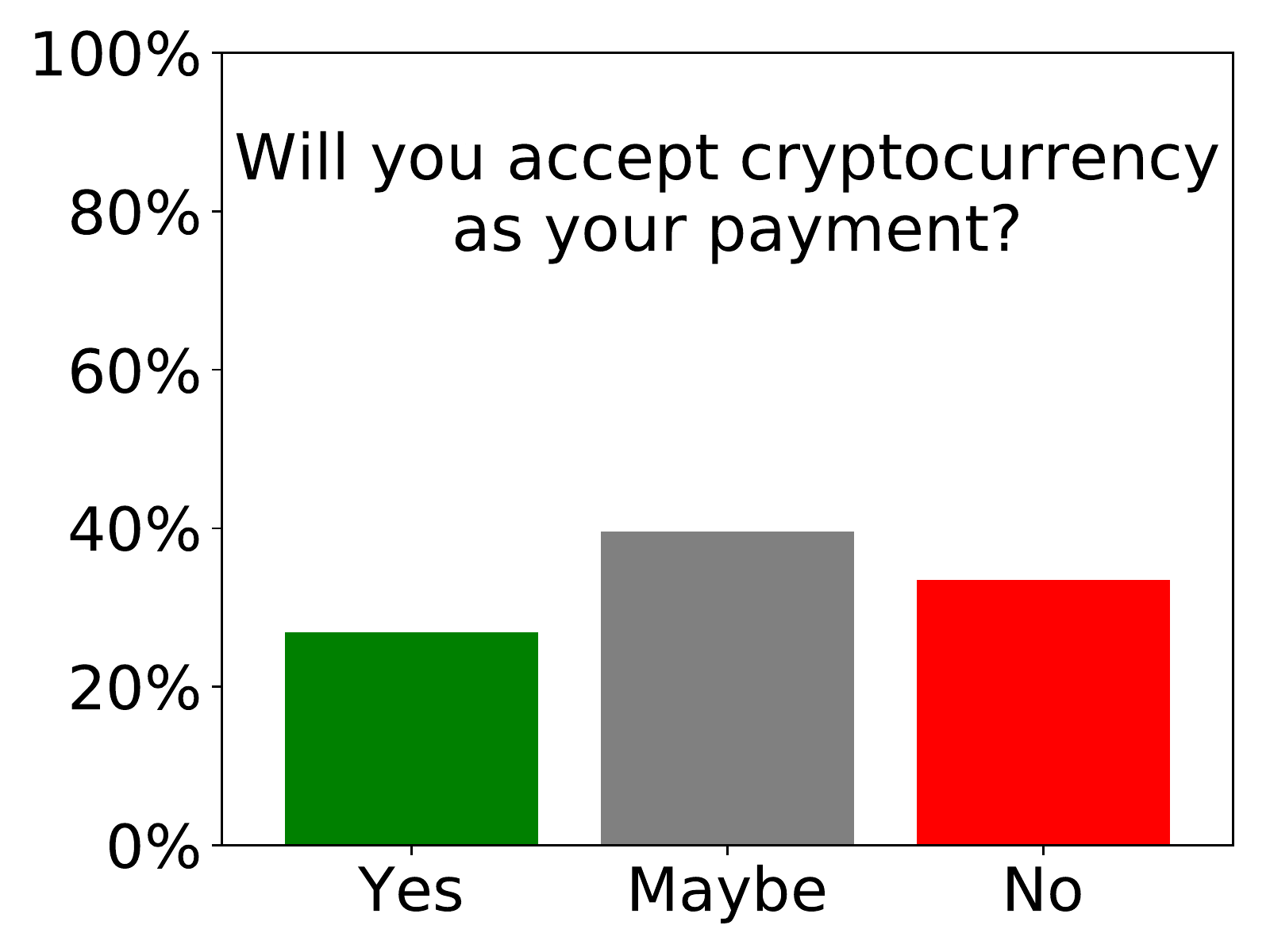}
        \vspace{-0.7cm}
        \caption{}
        \label{fig:key_finding:cryptocurrency}
    \end{subfigure}
    \vspace{-0.2cm}
    \caption{Key consumer survey findings.}
    \label{fig:key_finding}
\end{figure}

\begin{table}[]
\centering
\small
\begin{tabular}{|c|c|}
\hline
Concern & Fraction concerned \\
\hline
Security and Privacy & 82.8\% \\
Liability for Illegal Content & 50.5\% \\
Performance Impact on Device & 47.1\%\\
Payment Concerns & 42.1\% \\
Personal Ethics & 29.4\% \\
\hline
\end{tabular}
\caption{Concerns for users who specified that they would \textit{not} participate in the service.}
\label{t:participation_concern}
\end{table}

\begin{table}[]
\small
\centering
\begin{tabular}{|c|c|}
\hline
Concern & Fraction concerned \\
\hline
Don't Know How to Use/Sell & 59.7\% \\
Not Setup to Receive & 54.7\% \\
Volatility and Risk & 49.1\% \\
Don't Know What it is & 21.9\% \\
Other & 6.9\% \\
\hline
\end{tabular}
\caption{Concerns for users who specified that they would \textit{not} be willing to accept payment in cryptocurrency.}
\vspace{-5mm}

\label{t:cryptocurrency_concern}
\end{table}

\vspace{-3mm}
\para{Feasibility of participation.} To glean information about the potential for successful user participation,
we asked users about the devices they have to store/serve content, the free storage space of those devices, and the amount of time they would use those devices in the service.
83\% of users owned a laptop, and 82\% of laptop owners reported free storage space of more than 20 GB.
To put these numbers into context, consider that storing an average YouTube video that supports 5 bitrates requires 200 MB of space.\footnote{We obtained this number by scraping multiple reference videos with the youtube-dl tool~\cite{youtube-dl}. Reference videos were chosen to represent the average YouTube video length of 4 minutes~\cite{youtubestat}.} Peers with more than 20 GB of free storage can store more than 100 average videos. Further, consider that the average uplink capacity in the US is 22 Mbits/s~\cite{uploadspeed}. Streaming HD video requires an average of 5 Mbits/s throughput~\cite{netflixreco}, suggesting that peers should be able to stream up to 4 HD videos in parallel. Experiments with an Apache web server reveal that mean and peak CPU utilization (single core) are 0.36\% and 13\%, respectively, when serving 4 concurrent HD videos. These numbers mildly increase to 1.7\% and 33\% for 16 concurrent video streams.\footnote{Experiments were performed on a Desktop machine that has an Intel Xeon CPU with a 2.80 GHz processor.}

The remaining considerations are daily participation times and earnings for peers. 65\% of laptop owners in our survey stated that they use their computers for more than 2 hours a day. Serving content for 2 hours a day on a 22 Mbits/s link amounts to 580 GB of data served per month. If we assume that peers earn \$0.05/GB~\cite{cloudfrontprice}, they will make \$29 per month, which is greater than the requirement of covering half of their monthly Internet costs (the average US plan costs \$50 per month~\cite{broadbandprice}). We note that these numbers are conservative as 66\% of users would keep their laptops on for longer times, even when not in use, to serve (and earn) more.




\para{Concerns.} Figure~\ref{fig:key_finding:participation} shows that 49\% of users would not participate in the service. Table~\ref{t:participation_concern} lists the concerns shared by those users. As shown, a significant fraction of users were concerned about the impact that the service would have on their device, both with respect to security and privacy, and device/network performance. Thus, a practical deployment must ensure that service components running on user devices are sufficiently isolated from the rest of the host device, and are capped in terms of resources consumption. Many users would not participate due to content liability concerns, stemming equally from ethical considerations, legality, and privacy. Consequently, the majority of users were willing to serve movies/shows/news, but only 11\% were willing to serve adult content. These preferences promote distributed content filtering, and the inclusion of content information to the contracts between peers and content providers.


The other primary user concern was with respect to payments. 27\% of respondents were willing to accept payments in cryptocurrency, 40\% were unsure, and 33\% were against it (Figure~\ref{fig:key_finding:cryptocurrency}). Additional questions revealed that users were predominantly concerned with lack of familiarity with cryptocurrencies. Table~\ref{t:cryptocurrency_concern} shows that many users did not support cryptocurrency payment because they either did not know how to use/sell them, or they were not setup to receive them. Further, a significant fraction of users were concerned with the volatility and risk of cryptocurrencies. We expect these numbers to decrease in the upcoming years as the use of cryptocurrencies grows, and more advanced cryptocurrencies are created. However, these results do motivate the exploration of secure alternative payment forms.

\section{Secure and Practical Decentralized Incentivization}
\label{s:design}

In this section, we describe how content delivery and payments are handled in \name{}. We present solutions to the security vulnerabilities that arise from using decentralized monetary incentivization, and discuss several practical overhead considerations.

\subsection{Overview}
\label{ss:overview}

\name{} includes three major entities: a content provider that generates content (e.g., videos), 
a client who requests that content, and a peer who serves that content. To start, the content provider creates a \textit{Smart Contract} for each content file that they would like to distribute. The Smart Contract includes information about payments, describing how much peers will earn by serving this file, where payments will come from (i.e., the content providers cryptocurrency account), and the rules that peers must follow to prove that they served a file for payment (i.e., \textit{Proof of Delivery (PoD)}). Files are broken into chunks, and each chunk is placed on multiple peers who agree to the content provider's terms for that file.

\begin{figure}[]
    \centering
    \includegraphics[width=\columnwidth, height=1.6in]{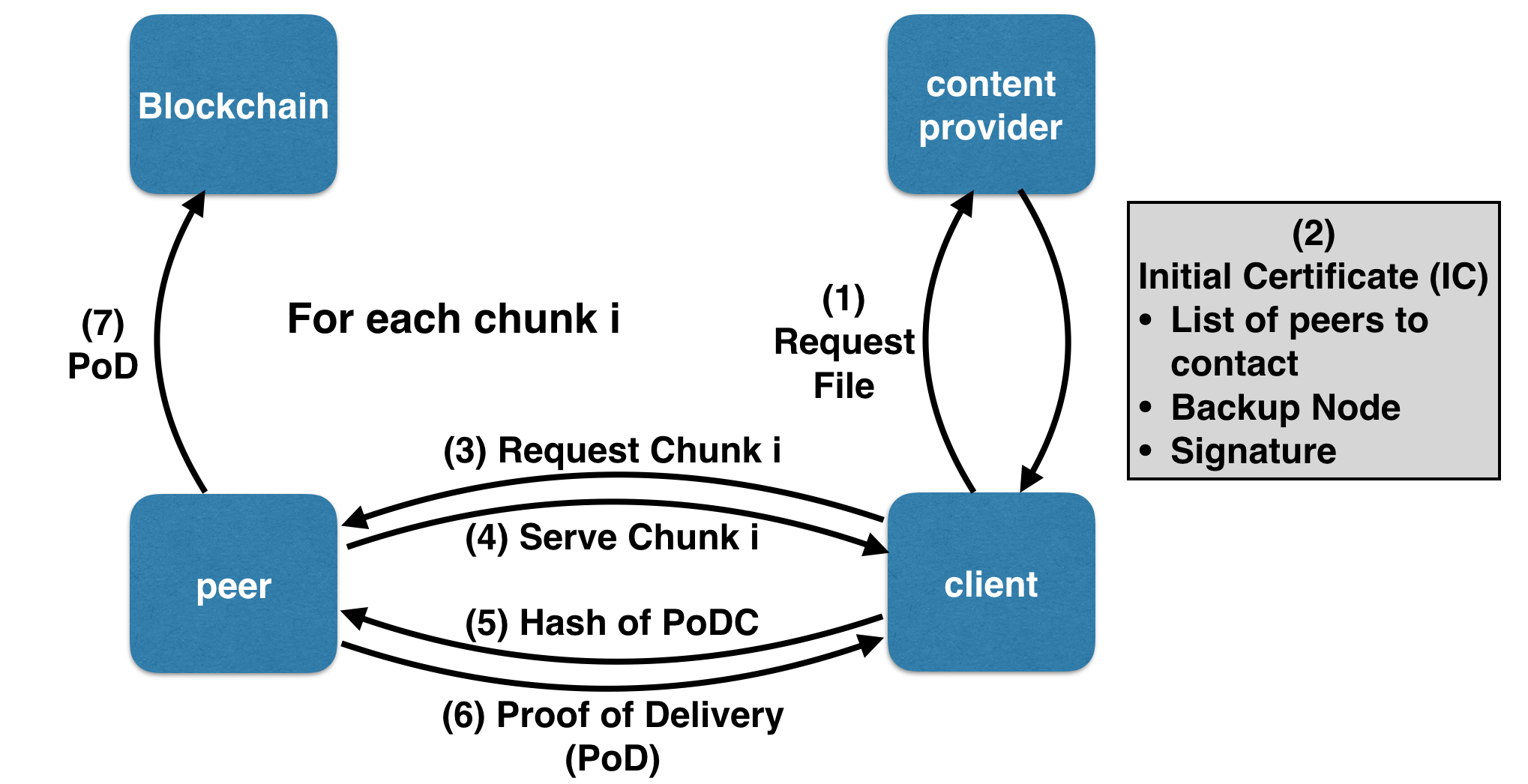}
    \caption{Downloading a file with \name{}.}
    \label{fig:request}
\end{figure}

Figure~\ref{fig:request} illustrates how \name{} handles a client request for a single file. Client requests are initially forwarded to the content provider, who responds with an \textit{Initial Certificate (IC)} that provides a guide on how to download all of the file's constituent chunks. The IC includes a list specifying the peer (identified by IP address and public key) to download each chunk from, along with the address of a Backup Node, which is a trusted server (e.g., a traditional CDN) to contact in the event that a peer is unreachable. Backup Nodes operate identically to normal peers, generating a PoD for each request that they serve.\footnote{To generate PoDs, Backup Nodes can be deployed on programmable CDNs like Amazon CloudFront~\cite{cloudfront}, which can execute arbitrary computations on each incoming request.} The content provider signs each IC with its private key to prevent forgery.

Upon receiving an IC, the client begins to download chunks from the listed peers. Clients maintain a sequential chain of PoD entries for each chunk they download, called the PoD chain (PoDC). The first element in the chain is the IC served by the content provider. After each successful chunk download, the client sends the peer a hash of the current PoDC. The peer then generates a new PoD (for the chunk they served) by signing the hash with its private key. The peer commits this PoD to the Blockchain, and also sends it to the client, which adds it to the PoDC. Thus, at the end of the file download, the Blockchain contains the entire PoDC for the file download. The PoDC effectively acts as a Blockchain, in that every link in the chain can be verified by a third party to ensure that the Smart Contract is not violated.

For simplicity, we described the client's downloading of chunks to be serial. This ensures sequential PoDs in the PoDC, which is necessary for third party verification. However, for improved performance, chunks for a file can be downloaded in parallel streams, such that each stream generates an independent PoDC comprised of chunks that are downloaded serially.



\subsection{Overheads}

\para{Content Provider Overheads.} With \name{}, every client file request is first forwarded to the content provider which generates an IC for the download. This overhead mimics that of a video download with a traditional CDN. Clients traditionally begin a video streaming session by first downloading a Manifest file directly from the content provider; video content is then downloaded directly from CDNs~\cite{pensieve}. In fact, with \name{}, content providers can simply append ICs directly to Manifest files. Manifest files can be further modified to include hash values for each chunk in the file, allowing clients to verify the integrity of the data they receive.

\para{Blockchain Overheads.} In the design presented in \S\ref{ss:overview}, each chunk download is recorded on the Blockchain by the corresponding peer. This transactional overhead may overwhelm the Blockchain. For example, 50,000 YouTube videos are watched every second, but existing Blockchains like Ethereum~\cite{ethereum} can only support 25 transactions per second.

To limit the overhead on the Blockchain, we propose using probabilistic payments. A simple approach is for content providers to grant payments only for a PoD which is divisible by a number N (specified in the Smart Contract). Peers would only submit a PoD to the Blockchain if this condition is met. Thus, reducing the number of Blockchain transactions by a factor of N.

One challenge with probabilistic payments is that the PoDC will no longer be directly recoverable from the Blockchain, since not all PoD entries will be present. However, the PoDC is necessary to verify that a PoD is valid, i.e., that it corresponds to a file delivery that adhered to the rules specified by the Smart Contract. To overcome this, peers can request the current PoDC from the client each time they generate a payment-eligible PoD. The peer can then include the PoDC in its Blockchain entry. It is important to note that PoDC transactions only occur once every N chunk downloads. Further, we expect the size of each PoD to be several bytes, implying PoDC sizes under a few KB.

Setting the value of N for probabilistic payments entails a tradeoff: a large N implies low payment frequency, while a low N limits the savings on Blockchain transaction costs. Assume that N is selected such than an active peer gets paid (on average) 10 times per day, or 300 times per month. We make two observations. First, given that existing Blockchains like Ethereum can handle 2.2 million transactions per day, \name{} would be able to support 220,000 active peers at any time. Second, payment frequencies are relatively stable. Specifically, the number of payments per month will be Binomial distributed, with a mean of 300 and a standard deviation of 17.3. Of course, the appropriate value of N will change over time, as new Blockchains are created. For example, Algorand~\cite{algorand} claims to support 100x more throughput (and thus users) than Ethereum.

\subsection{Detecting and Thwarting Attacks}
\label{ss:attacks}

Using monetary incentivization in decentralized settings has inherent security risks. For example, any involved entity (content provider, peer, or client) can be independently malicious, and entities can collude to perform even more complex attacks. In this section, we primarily focus on financially-motivated attacks, but we  also discuss other common attacks (e.g., DoS attacks).

\subsubsection{Single-entity Attacks}
\vspace{-2mm}

\para{Malicious peer.} To get paid, a peer must generate a PoD that is payment-eligible according to the Smart Contract. However, a malicious peer could attempt to generate a PoD \textit{without} serving content to a client. Preventing such an attack in \name{} is relatively straightforward. Recall that a peer generates a PoD using a hash of the PoDC that is provided by a client. If a malicious peer does not serve a chunk, the client can simply refuse to provide a PoDC hash (and subsequently download the chunk from the Backup Node). Further, since peers only learn the PoDC hash after serving a chunk, they cannot selectively choose to only service requests that will lead to a payment.

\para{Malicious content provider.} A content provider's goal is to maximize content distribution while minimizing cost. Since a content provider does not know a peer's private key (which is used to generated a PoD) in advance, it cannot predict what a PoD will look like with a new IC. However, a malicious content provider could generate ICs which have resulted in no payments to a peer during past downloads. To prevent this, ICs can be augmented to include a nonce, which serves as a unique identifier for each IC. Each content provider's nonces must be monotonically increasing (e.g., timestamps), and clients must include the appropriate nonce in each chunk request that they send a peer (i.e., Step 3 in Figure~\ref{fig:request}). In this way, peers can verify that incoming chunk requests are not intentionally designed to prevent payment by reusing a nonce value.

There is one complication with having peers verify that nonces are monotonically increasing. Given the decentralized nature of \name{}, client network latencies can create race conditions for nonce verification at peers. For example, consider a scenario in which two clients simultaneously request a file from a content provider such that the difference in their nonce values is one unit. The first client's network delay to load the first $n$ chunks could be greater than the corresponding delay for the second client, creating a scenario where the second client's request for chunk $n+1$ reaches the appropriate peer before that of the first client. To handle such scenarios, peers can maintain a sliding window of past nonce values. Incoming nonce values cannot match those in the window, and must be larger than the nonce value received immediately prior to the start of the window.


\subsubsection{Collusion Attacks}
\vspace{-2mm}

\para{Collusion between clients and content providers.} As noted above, a malicious content provider may want to prevent peers from receiving payments. While monotonically increasing nonces prevent the attack when a content provider acts alone, they are insufficient when content providers collude with clients. Specifically, a client (on behalf of a content provider) could fail to provide a PoDC to a peer that generates a payment-eligible PoD. Peers can certainly detect such behavior, and immediately stop serving content on behalf of the corresponding content provider. However, this poses a payment issue: peers receive probabilistic payments, so simply halting service for a content provider can yield significant amounts of wasted, uncompensated work. One way to mitigate this is for peers to request PoDC values from clients after every chunk they serve, rather than only doing so after they generate a payment-eligible PoD. In this way, peers would be able to quickly detect malicious clients. However, this solution adds significant bandwidth upload overheads to well-behaved clients.
Instead, peers should probabilistically request PoDC values from clients. If the probability of requesting a PoDC is higher than the probability that a peer receives a payment for a chunk download, peers can identify malicious behavior without wasting significant resources.


\para{Collusion between clients and peers (Sybil attack).} In an effort to increase payments, peers can collude with clients to lie about content delivery, thereby earning money without expending any resources. As a first step towards prevention, \name{}'s content providers randomize the peers that are responsible for serving the chunks of a file, making it difficult for a colluding client to contact a colluding peer. However, this is insufficient since the cost of creating a client is zero. A malicious peer can spawn a large number of clients to ensure that a significant number of clients contact it. In this model, each malicious client can generate a PoD for the first chunk in a file if it is intended to be served by a colluding peer; otherwise, the client can terminate the connection without wasting any resources.

Unfortunately, existing anomaly detection approaches to prevent collusion are unable to detect such attacks~\cite{rca}. Instead, \name{} enforces that the first chunk in each file is downloaded from a secure, trusted node (e.g., a traditional CDN). This modification creates an overhead to client generation, since a client \textit{must} expend bandwidth resources to download the first chunk of a file, before they can download subsequent chunks of the file from peers hoping to earn money. This overhead, in turn, makes Sybil attacks economically unfeasible.

To better understand this solution, consider a file download in which download bandwidth costs peers \$$x$ per chunk, content providers pay \$$y$ per chunk download (on average), and peer upload costs are \$$z$ per chunk. Further, assume that the fraction of malicious peers owning the chunks of that file is $m$, and the file comprises $l$ chunks. If peers and clients collude, the expected cost for a client to download the file up to chunk $i$ is,
\begin{align}
    \mathtt{Cost}\:[\:i\:] \;&=\; 1 \cdot x + \big(i-1\big) \cdot \big( m \cdot 0 + (1-m) \cdot x \big)
\end{align}
This equation states that the first chunk will be downloaded from a secure node (costing $x$). Additionally, if the remaining $i-1$ chunks
are routed to peers with equal probability, then a fraction $m$ of these chunks will be downloaded with zero cost, while the remaining chunks will be downloaded from honest peers. Similarly, the expected value of payments to a malicious peer for the file download until chunk $i$ is,
\begin{align}
    \mathtt{Payment}\:[\:i\:]  &= (i-1) \cdot m \cdot y
\end{align}

The attack can be deemed economically unfeasible if the cost to the colluding client is greater than the payments received by the malicious peer, or,
\begin{align}
    && \mathtt{Payment}\:[\:i\:]  \;&<\; \mathtt{Cost}\:[\:i\:] , & \forall i \in \mathbb{N} \nonumber\\ 
    \implies && y \;&\leq\; x \left( \frac{1}{m}-1 \right)
\end{align}

Further, to make the system monetarily feasible for peers, their compensation (\$$y$) should be greater than the bandwidth costs of uploading a chunk. This enforces an additional constraint
\begin{align}
    y \;&>\; z
\end{align}

These constraints bound payments based on the fraction of malicious peers serving the chunks in a file. For example, if the payment for a file is twice that of the download cost, then more than 33\% of the peers holding the chunks for that file must collude to make the attack economically feasible.


Can a malicious peer spawn a large number of colluding peers to achieve high values of $m$ for a particular file? Such an attack is unfeasible as a colluding peer will have to respond to requests from well-behaved clients, consuming peer bandwidth. Failing to respond to these requests can be easily detected by existing anomaly detection techniques. We do note, however, that the above approach works only if content providers ensure randomization of peer selection for chunk downloads \textit{and} content providers periodically mandate churn in the list of peers considered for a given file. Otherwise, a small number of malicious peers might still potentially achieve a high value for $m$.

Finally, a key property of the proposed solution is that it does not impose any overhead on well-behaved clients, as bandwidth is only used to download chunks in the requested file. However, requiring that the first chunk of each file is downloaded from a trusted node reduces the potential savings of using peer resources for content delivery, since infrastructure-backed nodes must now serve $1/l$ fraction of overall traffic.


\subsubsection{Non-financial Attacks}
\vspace{-2mm}
\para{Malicious client.} Though clients are not involved in the financial aspects of content delivery, they can still perform attacks on the system. Most notably, a group of malicious clients can perform a Denial of Service (DoS) attack on a content provider by failing to support payments to peers for serving that content provider's chunks. Peers that detect such behavior can decide to not serve the content provider's files, preventing content distribution. However, peers can only detect that a client is preventing payments after serving a chunk and requesting a PoDC from the client. A PoDC is requested with probability $p$ for each chunk download, implying that a client will have to download $1/p$ chunks from a peer before receiving a request for a PoDC. Since $p$ is intentionally kept to small values, the client will have to expend significant bandwidth resources (downloading chunks) in order to carry out such a DoS attack, making it impractical.

\section{Performance and Availability}
\label{s:perf}

Beyond secure incentivization, a decentralized CDN must perform numerous tasks to meet the requirements of content providers. In particular, providers expect their content to quickly be available to clients at all times. In this section, we outline how content providers can achieve these goals within \name{}.

\subsection{File Placement}
\label{ss:file_placement}
With \name{}, content providers must decide what peers serve their content. While the previous section discusses the security implications of file placement decisions, there are other factors which content providers will likely consider. In addition to supporting prior P2P file placement strategies~\cite{antfarm, iqiyi}, \name{} provides content providers with more flexibility in influencing content delivery pricing and performance.

\para{Availability.} Content providers must ensure that their files are available to their clients at all times, even in times of high demand. To do this, content providers must ensure that their files are hosted on a sufficient number of peers to tolerate bursts in request volume. Smart contracts provide a flexible way for content providers to influence the replication and availability of their files. Specifically, payment policies in Smart Contracts can be easily modified to reflect the current demands of a file. For example, a content provider can pay more for a file during peak demand times, akin to surge pricing with services like Uber.

\para{Quality of Service (QoS).} In addition to ensuring availability, content providers often aim to meet certain QoS goals for their client interactions. For example, many CDNs are judged based on the effective download speeds and latencies they provide~\cite{krishnamurthy2001use}. In order to give content provider control on QoS, each Blockchain record can be modified to include QoS metrics for the corresponding chunk download, as observed by the client. Additionally, peers can inform clients of the number of requests they are able to handle. With this information, content providers can factor in QoS decisions into their file placement and payment policies (via Smart Contracts). For example, a content provider can select a peer for a chunk download only if the predicted download speed from the peer is high enough to support the lowest bitrate for the video. We note that misreported QoS information by malicious clients can be easily detected (and mitigated) by taking into account all published QoS values for each peer.

To reduce per-chunk download delays, content providers should also incorporate geographical proximity of clients and peers in file placement decisions. For example, a popular file in a region warrants the content provider to insure sufficient peer representation of that file in the region. Content providers can leverage the flexibility of Smart Contracts for this as well, increasing payments to promote peers to serve files on their behalf.

\para{Resource Efficiency.} The resources that each peer provides are not fixed or identical. Thus, file placement strategies pose a complex resource allocation problem, particularly when multiple files and content providers are considered. Indeed, improper packing can lead to wasted peer storage/network resources. Chunk sizes are a knob which content providers can use to improve resource allocation. For example, smaller chunk sizes enable improved packing efficiency, and also reduce the network load on Backup Nodes that must serve the first chunk of each file. However, smaller chunks also increase transaction overheads on the Blockchain, and increase the bandwidth consumed by clients to send PoDC hashes to peers. This tradeoff implies that the appropriate chunk size will vary across content providers.


\subsection{Routing}
With \name{}, content providers send clients a list of peers to contact for each chunk in a file. Generating this list adds computational overhead to the content provider, who must keep track of the set of peers that is hosting its files, and which of those peers are alive at the time of a request. Detecting the set of live peers at any time is relatively straightforward~\cite{cohen2003incentives}, and incorrect listings can be handled by the Backup Node, ensuring availability.

The overheads of a centralized routing strategy motivate a decentralized approach. However, tools for decentralization, such as Distributed Hash Tables (DHTs)~\cite{chord,kademlia}, suffer from several challenges. First, randomizing peer selection with DHTs is difficult, since malicious nodes can directly influence routing decisions, enabling Sybil attacks~\cite{urdaneta2011survey}. Second, discovering the peer to contact for each chunk takes non-negligible time, harming client-perceived QoS. Third, in many scenarios, content providers will want to enforce routing policies that are QoS- and resource-aware (\S\ref{ss:file_placement}). Adding such information into a DHT is a challenge.
\section{Related Work}

\para{Secure Incentivization.} Past incentivization strategies for P2P file-sharing systems have focused primarily on preventing ``free-riding,'' where users consume resources but do not fairly contribute them back to the system~\cite{cohen2003incentives, qiu2004modeling}. Solutions have proposed using ``tit-for-tat'' mechanisms that require users to contribute resources in exchange for service. However, early approaches could not effectively prevent against malicious parties~\cite{bittyrant}. More recently, there have been proposals to use virtual currencies to solve the free-riding issue~\cite{karma,antfarm, dandelion}. In these solutions, users are granted and debited virtual currency for each transaction they participate in; a central authority is responsible for verifying all transactions. However, relying on a trusted central party is challenging (\S\ref{s:intro}). Further, unlike \name{}, all of these solutions incentivize only consumers of a service to become resource contributors, limiting adoption.




\para{Hybrid CDNs.} Many CDNs have attempted to supplement their traditional centralized infrastructure with cheaper resources contributed by peers~\cite{netsessions,iqiyi,livesky,kankan,tudou,spotify}. However, to date, these systems have struggled with peer adoption. Further, these approaches are not robust to attacks by malicious peers~\cite{rca}, and can thus benefit from the solutions presented in \name{}.


\para{Reliably Proving Service.} Content providers typically require that CDNs provide statistics about client interactions. As a result, several proposals exist for providing reliable accounting of client transactions in hybrid CDNs. Most notably, RCA~\cite{rca} requires clients to record download statistics in tamper-evident logs~\cite{peerreview}, which are later processed with anomaly detection algorithms. However, RCA is only able to detect attacks in which at least one of the two endpoints is honest (i.e., client or peer). Thus, unlike \name{}, RCA cannot handle the more complex collusion attacks (e.g., Sybil attacks) that arise with monetary incentivization.



\para{Peer-to-Peer Systems.} Numerous prior systems such as BitTorrent have motivated the potential of P2P file delivery services~\cite{freenet, kazaa, ripeanu2001peer,cohen2003incentives}. \name{} borrows promising techniques from these systems including policies on file placement and routing strategies (\S\ref{s:perf}). More recently, Filecoin~\cite{filecoin} 
is a P2P file storage service, which uses cyptocurrency-based incentivization to attract peers to store files. In contrast to \name{} which uses a ``Proof of Delivery'' mechanism to verify transactions, Filecoin relies on a ``Proof of Storage'' technique which cannot be directly applied to content delivery. Moreover, Filecoin does not let third parties (e.g., content providers) sponsor a client's download of a file.



\section{Discussion and Future Work}
\label{s:discussion}

\para{Conflicts with ISPs.} \name{}'s use of peer resources for content delivery presents a tension with the goals of ISPs. In particular, having peers serve files increases the amount of upstream traffic, and correspondingly costs, for ISPs. However, prior work has demonstrated that this tension can be resolved with intelligent peer selection algorithms, in which peers within the same autonomous system as the client are preferentially selected~\cite{taming,netsessions}. Such a policy can significantly reduce inter-ISP traffic, mitigating the cost effects of peer-to-peer content delivery.

\para{Using \name{}.} In order to use \name{}'s content delivery service, content providers must modify their websites or applications to adhere to the protocols described in Section~\ref{s:design}. For example, web pages can use a JavaScript library to request ICs, contact peers for chunks, and maintain (and compute hashes of) PoDCs. This approach, in contrast to having a cross-application shim layer running on a client's machine, gives content providers the option to optimize for different metrics (\S\ref{s:perf}), and obviates the security challenges associated with shared proxy solutions.

\para{Widespread Peer Adoption.} In this paper, we considered peers to be personal computers (i.e., laptops or desktops). However, in practice, a plethora of other devices could serve as peers, improving performance and benefits. For example, in home settings, continually-powered WiFi routers or IoT devices (e.g., security systems) could serve as reliable peers. On the opposite end of the spectrum, colocation datacenters might deploy unused machines as peers. \name{}'s protocols for secure, decentralized content delivery generalize to peers with diverse resource profiles.

\para{Generalized Resource Sharing.} \name{} introduces several new ideas to support decentralized, secure incentivization in content delivery settings. However, can these ideas be applied to more generalized resource sharing settings? A key challenge to generalization is creating a policy which can prove that a service-specific task was performed correctly. For example, \name{} requires ``Proof of Delivery,'' while Filecoin~\cite{filecoin} uses ``Proof of Storage.'' Similarly, clients may want to offload GPU-based tasks to peers rather than to costly cloud services; such a service would require ``Proof of Faithful Computation.''

\bibliographystyle{abbrv} 
\begin{small}
\bibliography{marlin}
\end{small}

\end{sloppypar}

\end{document}